\documentclass{article}
\usepackage[T1]{fontenc}
\usepackage[utf8]{inputenc}
\usepackage[portrait]{geometry} 

\usepackage{authblk}
\usepackage{amsmath,amssymb,amsthm,amsfonts}
\usepackage{esint}
\usepackage{empheq}
\usepackage{xspace}
\usepackage{enumitem}
\usepackage{hyperref}

\usepackage[backend=biber,style=apa,citestyle=apa]{biblatex}
\setlist{nosep}

\newcommand{\ww}{{\small{WAVEWATCH III}}\textsuperscript{\tiny\textregistered}}

\newcommand{\wam}{{\small{WAM}}}
\newcommand{\era}{{\small{ERA5}}}
\newcommand{\eraint}{{\small{ERA5-Interim}}}
\newcommand{\waverys}{{\small{WAVERYS}}}
\newcommand{\glorys}{{\small{GLORYS12}}}
\newcommand{\ecmwf}{{\small{ECMWF}}}
\newcommand{\cmems}{{\small{CMEMS}}}
\newcommand{\gebco}{{\small{GEBCO}}}
\newcommand{\emodnet}{{\small{EMODNET}}}
\newcommand{\etopo}{{\small{ETOPO1}}}

\newcommand{\punderline}[1]{\underline{\smash{#1}}}

\newcommand{\dpar}[2]{\frac{\partial {#1}} {\partial {#2}}}
\newcommand{\dcpar}[2]{\frac{\partial^2 {#1}} {\partial {#2}^2}}

\DeclareMathOperator{\pgrad}{\punderline{\nabla}}

\numberwithin{equation}{section}

\begin{document}

\title{Current interaction in large-scale wave models with an application to Ireland}
\author[1,2]{Clément Calvino}
\author[2]{Tomasz Dabrowski}
\author[1]{Frederic Dias}
\affil[1]{University College Dublin, Dublin, Ireland}
\affil[2]{Marine Institute, Rinville, Ireland}
\date{}

\maketitle

\begin{abstract}
    A nested large-scale wave model that covers most of the Atlantic Ocean and focuses on generating accurate swell conditions for Ireland is presented and validated over the two years $2016\mathrm{-}2017$.
    The impact of currents is studied using the surface currents from the \glorys{} product. Currents slightly reduce by $1\%$ the error of the model compared to altimetry data, but they explain most of the wave energy at scales less than $50 \,\mathrm{m}$.
    The variability induced by currents is found to be more noticeable on the instantaneous fields, up to $50 \,\mathrm{cm}$. Track following observations indicate that wave refraction induced by mesoscale eddies is correctly captured. However, this modulation is of the same order as a shorter-scale variability appearing in the altimetry data, which makes it difficult to objectively assess the impact of currents.
\end{abstract}

\section{Introduction}

Traditionally, large-scale simulations or hindcasts have been focusing on the air-sea interaction, trying to balance correctly wind input and swell dissipation. More recently, there has been some interest in the interaction with currents, which for the most part highlights the refraction induced by current gradients. The impact of large-scale currents like the Agulhas current or the Gulf Stream is well documented through numerical simulations cross-checked with satellite observations - see for example \cite{marechal2020surface}; \cite{barnestowards} for recent studies on the Agulhas current. The impact of small scale currents has drawn less attention but in \cite{ardhuin2017small} it is shown that even submesoscale structures with shorter length scales (ranging from $10$ km to $100$ km) can explain the majority of the significant wave height spatial variability. 

In wave spectra models, the interaction of currents on wave propagation appears in the wave action balance equation as a correction of the wave group velocity by a current velocity term. The framework used to derive this equation is usually a depth-uniform current, see for instance \cite{peregrine1976interaction}. However, in the real world, the current is rarely uniform with depth. Therefore approximations are made, which differ depending on the application. A few are listed below.

Global wave models are computationally expensive. For this reason the current interaction is often ignored as the wind input and the bathymetry are the most important factors - see for example \cite{stopa2018wind}. In that paper, the wind is calibrated to the wave growth parameter within the spectral wave model \ww{} for 10 reanalysis datasets and 2 datasets composed of merged satellite observations. It is demonstrated that the space-time distributions of extreme waves are very different even after calibration. But wave-current interaction is ignored. 
The European Centre for Medium-Range Weather Forecasts (\ecmwf{}) and the Copernicus Marine Environment Monitoring Service (\cmems{}) each run a different forecast wave model based on the Wave Model numerical code (\wam{}). None takes currents into account. However both the \ecmwf{} and \cmems{} provide a wave reanalysis (\era{} and \waverys{} respectively) coupled with an ocean model that takes into account the effects of currents on wave propagation.
Large-scale wave models can have a more refined grid where the impact of current is more significant. In \cite{marechal2020surface}, \cite{ardhuin2017small}, and \cite{barnestowards} an offline coupling is done using an external current data-set. All these authors are directly interested in the wave height spatial variability caused by ocean currents. Different products are used for the currents but in each case they feature the surface current velocity. It is a cost efficient choice and still a good approximation in deep water (\cite{kirby1989surface}; \cite{banihashemi2017approximation}). 
In a similar fashion the \cmems{} runs a forecast model for the European North-West Atlantic shelf using \ww{} forced among others by surface ocean currents from one of their forecast ocean models.

The work presented in this paper consists of a large-scale application making use of an offline coupling to compute the current induced interaction terms. A standalone \ww{} model is set-up for the Atlantic and the resolution is refined around Ireland with two added nested grids. A sensitivity analysis on the wind input parameters is carried out first. It is shown that there is no single way to optimise the wind input growth term as the distribution does not react homogeneously to the growth parameter. The current interaction term in the wave action balance equation is computed. The current field corresponds to the surface currents from the 3D global ocean reanalysis \glorys{} provided by the \cmems{}. The impact of currents is evaluated by comparing the output of the model against satellite altimetry and wave buoy data, focusing on the North-East Atlantic region around Ireland. It is shown that currents can locally have a noticeable effect on wave propagation. However, they marginally impact the accuracy of the model on average. In Section 2, the wave model is described. Section 3 explores the sensitivity to wind input. The impact of the current field is investigated in Section 4. Conclusions are provided in Section 5.

\section{Description of the wave model}

A large-scale standalone \ww{} model (v6.07) is set-up and the different choices for the grid and model parameters are listed below.
The wave action balance that is usually implemented in the most recent third generation wave models, like \ww{}, assumes a vertically uniform current that can slowly vary in the horizontal directions (the short-wave or large-scale current approximation). A complete derivation is given in \cite{peregrine1976interaction} and the same type of derivation is found in the more recent literature like \cite{holthuijsen2010waves}.

The wave variables are denoted without a prime while the relative wave variables are denoted with a prime, except for the frequency where $\sigma$ is the intrinsic frequency and $\omega$ the absolute frequency, related through the following dispersion relation highlighting a Doppler shift caused by the currents:

\begin{empheq}{align}
 & \omega = \sigma + \punderline{U}\cdot\punderline{k} \,. \label{doppler}
\end{empheq}

The following definitions are used for the absolute wave phase speed $c$ and wave group velocity $\punderline{c}_g=(c_{g,x},c_{g,y})$. Their intrinsic counterparts are derived by using the dispersion relation, with $\punderline{U}=(U_x,U_y)$ the current field passed down to the wave model and $\punderline{k}=(k_x,k_y)$ the wave vector with $k$ its modulus:

\begin{empheq}{align}
 & c = \frac{\omega}{k} = \frac{\sigma + \punderline{k}\cdot\punderline{U}}{k} \,, \label{phase} \\
 & \punderline{c}_g = \pgrad_{\punderline{k}}{\omega} = \pgrad_{\punderline{k}}{\sigma} + \punderline{U} = \punderline{c}_g' + \punderline{U} \,. \label{group}
\end{empheq}

The wave action balance is given by Eq. (\ref{wa}) below, with $A(x,y,t, k,\theta)$ the wave action, $(x,y)$ the eastward and northward horizontal coordinates, $t$ the time, $\sigma$ the intrinsic wave frequency, $\theta$ the spectral direction, $c_{\theta}$ and $c_{k}$ the rates of change of the direction $\theta$ and wavenumber $k$ respectively, $h$ the mean water depth, and $S$ a global source term:

\begin{empheq}{align}
 & \dpar{A}{t} + \dpar{(c_{g,x} A)}{x} + \dpar{(c_{g,y} A)}{y} + \dpar{(c_{k}A)}{k} + \dpar{(c_\theta A)}{\theta} = \frac{S}{\sigma} \,, \label{wa} \\
 & \punderline{c}_g = \punderline{U} + \punderline{c}_g' = \punderline{U} + \frac{\punderline{k}}{2k} \left(1 + \frac{2kh}{\sinh{2kh}}\right) \sqrt{\frac{g}{k} \tanh{kh}} \label{eq:3} \,, \\
 & c_{k} = \frac{D k}{D t} =  - \frac{1}{k} \dpar{\sigma}{h} \bigg[ \punderline{k} \cdot \left(\pgrad{h}\right) \bigg] -  \frac{\punderline{k}}{k} \cdot \bigg[ \punderline{k} \cdot \left(\pgrad \punderline{U}\right) \bigg] \,, \\
 & c_{\theta} = \frac{D \theta}{D t} = - \frac{1}{k^2} \dpar{\sigma}{h} \bigg[ \dpar{\punderline{k}}{\theta} \cdot \left(\pgrad{h}\right) \bigg] - \frac{1}{k^2} \dpar{\punderline{k}}{\theta} \cdot \bigg[ \punderline{k} \cdot \left(\pgrad{\punderline{U}}\right) \bigg] \,.
\end{empheq}

\ww{} (\cite{wavewatch2019user}) is a widely used third-generation wave model accounting for a wide range of processes. The source terms used in this paper are gathered in $S$ (see Eq. (\ref{wa})) and discussed briefly in the next paragraph, but a more exhaustive list is available in the \ww{} manual. It was shown to give excellent results for global simulations and is used in many operational models or reanalysis products (\cite{ardhuin2011calibration}; \cite{rascle2013global}; \cite{stopa2018wind}). Different parameterizations are offered either for the propagation terms (left-hand side of Eq. (\ref{wa})) or for the source terms (right-hand side of Eq. (\ref{wa})). For the present application the source terms can be written as in Eq. (\ref{sources}) and some impactful choices are explicitly detailed in the next section:

\begin{empheq}{align}
 & S = S_{\text{ln}} + S_{\text{nl}} + S_{\text{in}} + S_{\text{ds}} + S_{\text{bot}} + S_{\text{db}} \,, \label{sources}
\end{empheq}

where the terms denote, respectively, linear growth, nonlinear transfer of wave energy through three-wave and four-wave interactions, wave growth by the wind, wave decay due to whitecapping, bottom friction and depth-induced wave breaking.

\subsection{Current effects}

The equations introduced above can only be derived for a vertically uniform current. Assuming a more realistic sheared current, these equations can still be used but the current field $\punderline{U}$ becomes an effective current field, which in reality is a higher-order correction for the wave group velocity as explained in \cite{kirby1989surface} and \cite{banihashemi2017approximation}. For this paper, the surface currents are used, which is common practice for large-scale wave models in deep water.

As shown in Eqs (\ref{wa})--(\ref{doppler}), currents are impacting wave propagation at different levels, and they can also impact the source terms in $S$. In \cite{ardhuin2012numerical}, the main current impacts are shown to be the current-induced refraction through $c_{\theta}$, and the relative wind effect where the wind speed is taken relative to the currents in the wind input source term.
It is worthwhile to note that similarly to the effect of current-induced refraction through $c_{\theta}$, the other current effects (current advection and Doppler shift) also contribute to increase the wave energy, and therefore wave heights, in areas of stronger opposing currents.

\subsection{Grid set-up and parametrization}

\begin{figure}
  \centering \includegraphics[width=15cm]{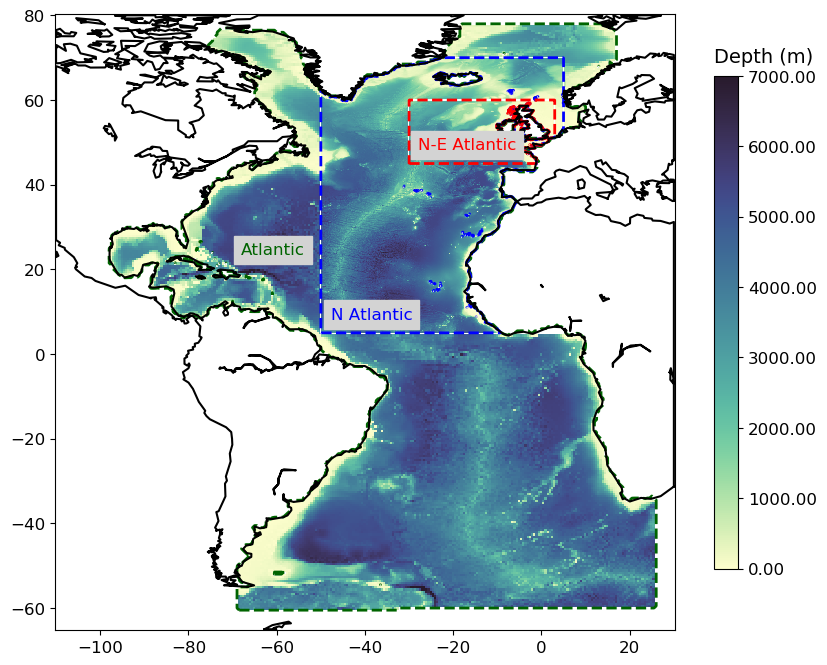}
  \caption{The three different grids used in the \ww{} model drawn on top of the \gebco{} bathymetry. Their specifications are given in table \ref{grids1}. The North-East Atlantic grid includes the Irish shelf where shallower waters are found. The North Atlantic grid is ensuring an acceptable parent-to-child resolution ratio ratio between the grids.}
  \label{grids2}
\end{figure}

Three regular grids are used. They become more and more refined as one gets closer to Ireland, which is the area of interest for this study. The goal is to capture all the storms possibly impacting the wave conditions in Ireland, so the model does not require wave boundary conditions but only wind input. The grid specifications are summarized in table \ref{grids1} with a graphical representation given in figure \ref{grids2}. The same common values are used for the three associated spectral grids and the spatial ratio between a child and a parent grid does not exceed $5$, which is a commonly accepted value. The grid configuration is relatively similar to a previous wave hindcast focusing on the North-East Atlantic Ocean (\cite{pilar200844}).

\begin{table}[h]
\begin{center}
\begin{tabular}{l|c|c|c}
  Grid & Atlantic & North Atlantic & North-East Atlantic \\
  \hline
  Grid points & $282 \times 292$ & $552 \times 652$ & $662 \times 302$ \\
  \hline
  Resolution (degrees) & $0.5$ & $0.1$ & $0.05$ \\
  \hline
  Global time-step (s) & $1800$ & $600$ & $300$ \\
  \hline
  \hline
  First frequency (Hz) & Frequency increment & Frequency bins & Direction bins \\
  \hline
  $0.0373$             & $1.1$               & $36$           & $36$ 
\end{tabular}
\caption{Grid specifications used in the \ww{} model. The goal is to obtain the best resolution possible for the North-East Atlantic grid, while conserving a similar number of grid points between all grids to balance the computational load.}
\label{grids1}
\end{center}
\end{table}

The source terms that are included in this paper are highlighted in Eq. (\ref{sources}). For the most part they consist of standard choices suited for a large-scale application. The linear growth term $S_{\text{ln}}$ is parameterized as in \cite{cavaleri1981wind}. It is useful to make the wave field grow from calm conditions. The nonlinear wave-wave interactions $S_{\text{nl}}$ are modeled with the Discrete Interaction Approximation from \cite{hasselmann1985computations}. The source package ST4 (\cite{ardhuin2010semiempirical}) is used. It encompasses the wind input $S_{\text{in}}$ and wave dissipation $S_{\text{ds}}$. The impact of bed roughness appears through the bottom friction term $S_{\text{bot}}$ with the parametrisation of \cite{grant1979combined}, and the depth-induced breaking $S_{\text{db}}$ is parameterized using \cite{battjes1978energy}.

\subsection{Wind input parametrization}

A first sensitivity analysis is performed to adjust the wind input parametrization. Default values are provided in \cite{ardhuin2010semiempirical}, but it is always recommended to adjust the most impactful parameters for a specific application (\cite{ardhuin2011calibration,stopa2018wind}). The choice of parameters for the wind input notably depends on the wind input source used and on the region for which the model is validated. The error on the output is not spatially homogeneous and different biases can be observed for different parts of the ocean (\cite{ardhuin2011calibration}). These biases can be locally corrected by adjusting the model parameters.

The wind input source term is part of the source package ST4 (\cite{ardhuin2010semiempirical}) that also encompasses dissipation by whitecapping, which is an energy flux from the wave field to the ocean currents due to wave breaking not induced by the depth, and swell dissipation, which is a flux of energy from the wave field to the wind. Going quickly through the theory of wind-wave growth and referring here to the textbook \cite{ardhuin2020ocean}, we emphasize that if the initial growth of waves is well explained by the turbulence theory developed in \cite{phillips1957generation}, the further growth of waves comes from the feedback of the waves on the atmosphere. The wave-induced pressure is roughly in quadrature with the surface elevation and that phase shift enables the transfer of energy from the air to the wave field at the surface. Using the linear equations from Airy's wave theory and adding an atmospheric term in quadrature with the surface elevation gives the simplified evolution of the wave amplitude $a$, shown below in Eq. (\ref{growth1}), where $\beta$ is a growth parameter. The right-hand side is then generalized for a wave system giving the spectral wind input source in Eq. (\ref{growth2}), with $N = \sigma A$ the wave energy spectrum:

\begin{empheq}{align}
 & \frac{d}{d t} \left(\frac{a^2}{2}\right) = \sigma \beta \frac{a^2}{2} \label{growth1} \,, \\
 & S_{\text{in}} (k,\theta) = \sigma \beta N(k,\theta) \,. \label{growth2}
\end{empheq}

The $\beta$ term in Eq. (\ref{growth2}) can be modeled by different formulations. Several options are available in \ww{} through the choice of the source package, and those parametrizations keep being refined as more and more field observations, experiments and high-resolution numerical simulations are made. The source package ST4 is taking into account several new features. For instance it is observed that the interaction between the air flow and the waves is stronger for younger waves, and at the same time the detachment of the air flow occurring for high winds is decreasing the wave growth, which can explain why the drag coefficient is observed to be reduced during hurricanes. A sheltering effect from the long waves on the short waves is also expected to reduce the growth of short waves. The wind input is computed as shown in Eq. (\ref{input1}) with $\rho_a$ and $\rho_w$ the air and water densities, $\kappa$ the von K\'arm\'an constant, $c = \omega/k$ the wave phase speed, $U_r$ and $\theta_r$ the reference wind velocity and direction input at the height $z_r$. The wind velocity $U_r$ is the relative wind speed velocity, corrected by the current. It is adapted from \cite{janssen1991quasi} with a correction for the friction velocity as in \cite{chen2000effects} given by Eq. (\ref{input2}) to include the sheltering effect of long waves on short waves, with $s_u$ a tuning sheltering coefficient ranging from $0$ to $1$:

\begin{empheq}{align}
 & S_{\text{in}} (k,\theta) = \sigma \frac{\rho_a}{\rho_w} \frac{\beta_{max}}{\kappa^2} e^Z Z^4 \frac{v_*^2 (k)}{c^2} \max(\cos(\theta-\theta_x),0)^2 N(k,\theta) \,, \label{input1} \\
 & v_*^2 (k)= u_*^2 - |s_u| \int_{0}^k \int_{0}^{2\pi} \frac{S_{in}(k',\theta)}{c} \, dk' d\theta \,. \label{input2}
\end{empheq}

The wave age $Z$ is given by Eq. (\ref{age}) with $z_\alpha$ a wave age tuning parameter:

\begin{empheq}{align}
& Z = \log(kz_1) + \kappa / \left( \cos(\theta - \theta_u)(u_*/c + z_\alpha) \right) \,. \label{age}
\end{empheq}

The roughness height $z_1$ is evaluated from the following set of equations (\ref{z1}):

\begin{empheq}{align}
 & U_{r} = \frac{u_*}{\kappa} \ln\left(\frac{z_{r}}{z_1}\right) \,, && z_1 = \frac{z_0}{\sqrt{1 - \tau_w/\tau}} \,, && z_0 = \min\left(\alpha_0 \frac{\tau}{g},z_{0,max}\right) \,, && \tau = u_*^2 \,. \label{z1}
\end{empheq}

The friction velocity $u_*$ is evaluated with a law of the wall from the reference wind velocity $U_{r}$ known at the height $z_r$. The roughness height then depends on the wave stress $\tau_w$ which is evaluated by Eq. (\ref{stress}) from \cite{janssen1991quasi}, and $\tau$ is the total stress related to the friction velocity. The roughness height is evaluated by taking into account the effect of high-frequency waves with $z_0$ given by the Charnock relation and $\alpha_0$ the Charnock coefficient. It is capped with a user defined value $z_{0,max}$ to avoid unrealistic high values in extreme high wind conditions. In Eq. (\ref{stress}), $c$ and $c_g$ are the phase speed and wave group velocity modulus, and $\chi$ is the normalized vertical component of the wave-induced velocity in the air, found to be related to the stress tensors:

\begin{empheq}{align}
 & \tau_w = - \int_0^\infty D_w \dcpar{U_{x}}{z} \, dz \,, && D_w = \pi \int_0^{2\pi} \frac{\sigma^2 k}{|c - c_g|} |\chi|^2 N(k,\theta) \cos^2(\theta) \,d\theta \,, && \chi = \tau_w / \tau \,. \label{stress}
\end{empheq}

Several tunable parameters are appearing in the parametrization of the wind input described above but the $\beta_{max}$ parameter is the only parameter globally and homogeneously affecting the wind input source term. It is directly controlling the amount of energy put into the model from the wind input with no distinction on the wind speed, wave direction or frequency. It is controlling the overall wave growth rate and significant wave height bias in the model. Different sets of parameters are offered by default in \ww{}. They were found after a meticulous spectral analysis of the different source terms involved in the package (\cite{ardhuin2010semiempirical}), reaching a balance between the dissipative terms and the growth terms in key parts of the spectrum notably. However slightly different input and forcing data-sets are used here, bearing different biases. Adjusting the parametrisation is therefore a necessity. As done in \cite{stopa2018wind}, we decided to focus only on the growth rate $\beta_{max}$.

\subsection{Input fields}

The different input fields that we used are described in this section. The spatial and temporal resolutions of each product are highlighted. The way the data is processed when applicable is also briefly mentioned.

\subsubsection{Bathymetry: \gebco{}}

The latest version of \gebco{} is used (\cite{tozer2019global}). It is a processed product offering a global bathymetry data on a regular $1/4$ arc minute resolution grid, that gathers and merges different sources together. Most of the data for Europe comes from the processed product \emodnet{} (\cite{emodnet2018emodnet}), which is itself a merged product gathering local surveys from each of the partners in the consortium with a final resolution of $1/16$ arc minute. The gaps in the bathymetry are filled with \etopo{} (\cite{amante1arc}), which is a global $1$ arc minute resolution processed product using satellite altimetry data.

The bathymetry from \gebco{} is then interpolated with a classic bi-linear scheme on each computational grid described previously. The most refined grid is the North-East Atlantic grid with a $3$ arc minute resolution, which is a down-grade in resolution compared to \gebco{}, meaning the bathymetric product we use is well suited for the wave model presented in this paper.

\subsubsection{Wind input: \era{}}

The wind growth is the main driving mechanism ruling the generation of waves in our case. As such the wind input data used for the model is the most sensitive input data and its quality and resolution are impacting the accuracy of the model more than the bathymetry or the current field. The \era{} data-set (\cite{hersbach2020era5}) is a reanalysis offered by \ecmwf{} using a 4D-Var data assimilation scheme on one of their forecast systems, covering the period from $1979$ onwards. It provides hourly time series for a variety of atmospheric variables on a regular $25$ arc minute horizontal resolution grid with $137$ hybrid sigma/pressure (model) levels in the vertical. The horizontal resolution of the wind input is well suited to generate the swell in the coarser global grid covering the Atlantic ($30$ arc minute resolution). However the two refined grids used for the present application have a higher resolution than the \era{} grid (respectively $6$ and $3$ arc minute). Therefore there is a loss of spatial accuracy due to the resolution of the wind input. More precisely, scales lower than $15$ arc minute resolution spatially (around $17$ km at a 53N latitude) are not resolved by the wind input data, and therefore are not resolved by the wave model either. The interpolation of the forcing data is done on the fly by \ww{} with a quadratic interpolation in time and a bi-linear interpolation in space.

The wind speed is also adjusted with the current velocity provided to the wave model. It was shown in \cite{bye1999atmosphere} and more recently in \cite{renault2016modulation} that it can significantly impact the momentum transfer from the atmosphere to the water column. In \cite{ardhuin2012numerical}, a more thorough analysis was conducted that looked specifically at the impact of relative wind speed. A better agreement was found when correcting the $10\,\mathrm{m}$ wind speed with the ocean surface current. It was also mentioned that using only a fraction of surface current could be a better choice although it was not specifically tested. The literature suggests a clear feedback from the ocean current on the atmospheric boundary layer, surface wind stress and consequently the wave action wind input and dissipation. It is however still unclear what are the exact parameters influencing this interaction and as a result numerical studies seem to have made arbitrary and empirical choices in their parametrization. In our case we are using the surface currents.

The \era{} data is showing a negative bias for high winds (\cite{ardhuin2011calibration}; \cite{rascle2013global}). This bias can be partly corrected by tuning adequately the parameters for the wind input, but a more straightforward approach is to directly increase the wind velocity for high winds. This is done by applying the correction (\ref{windcor}) below, with $W$ the wind speed and $W_t$, $c_t$ two constants that can be tuned to set the threshold where the correction starts to apply. This correction is available in \ww{}. The values used in the present application have been suggested by Ardhuin's team in Ifremer:

\begin{empheq}{align}
 & \mathrm{If} \quad (W > W_t), \quad W = W + (W-W_t)c_t, \qquad W_t = 23 \,\mathrm{m.s^{-1}}, \quad c_t = 1.08 \label{windcor} \,.
\end{empheq}

\subsubsection{Current: \glorys{}}

The current and sea level fields are of second importance in most large-scale wave models. It is the purpose of this paper to evaluate how sensitive the model is towards these fields. The global ocean eddy-resolving reanalysis \glorys{} delivered by Mercator Ocean as part of the \cmems{} (\cite{PHY_001_030}) is used for these two fields. It covers the years from $1993$ to $2019$ and is forced using atmospheric data from the \eraint{} reanalysis (\cite{dee2011era}). It is assimilating along track altimetry data and gives daily averaged time series of the main oceanic variables of which only the horizontal velocities and surface elevation are used for the present application. The data is provided on a global regular $5$ arc minute resolution grid with $50$ vertical levels well refined at the surface to allow for a consistent representation of the different air-sea processes. The horizontal resolution is comparable with the level of refinement used in the computational grids of this model, but given the time average only the long-term components of the currents and surface elevation are consistently retained. As mentioned in \cite{lellouche2018recent}, the submesoscales are not captured, ruling out most of the variability in storms, and the strongest component observed in the data is the geostrophic component. Tides are also not included in this model.

The first layer of the \glorys{} data is used at $0.5 \,\mathrm{m}$ depth, for the surface currents. The current and surface elevation fields are then interpolated on the fly by \ww{} with a linear interpolation in both time and space.

\section{Wind input sensitivity}

The \ww{} model described previously is run for the years $2016$ and $2017$. Five values for the $\beta_{max}$ parameter introduced in Eq. (\ref{input1}) are tested, ranging from $1.60$ to $1.80$. The default value of $\beta_{max}$ for \ecmwf{} winds is $1.43$, and the maximum value tested and mentioned in the manual is $1.55$. However, in \cite{stopa2018wind}, the growth rate is tuned for different wind input sources used in a global wave model application. Values between $1.10$ and $2.05$ are found. In order to validate the model, statistics are computed, averaging over the full duration of the simulation but excluding the first month to avoid unnatural errors due to the model spinning up. The validation is conducted by comparing the model output to wave buoy and weather buoy data, altimetry data and then comparing the accuracy of the model with a reference wave hindcast simulation done for Ireland by \cite{gallagher2014long}.

The following statistics are used for the validation using the generic notations $X$ and $Y$ for the time series with $N$ records. When the output of the model is compared against an observation, the model data is interpolated in time and space to match exactly the observation:

\begin{empheq}{align}
 & \mathrm{mean} && \mathrm{m}(X) = \frac{\sum_{i=1}^{N} {X_i}}{N} \,,\\
 & \mathrm{standard\,deviation} &&  \sigma(X) = \sqrt{\frac{\sum_{i=1}^{N} (X_i - \mathrm{m}(X))^{2}}{N}} \,,\\
 & \mathrm{root\text{-}mean\text{-}square\,error} && \mathrm{rmse}(X,Y) = \sqrt{\frac{\sum_{i=1}^{N} (X_i - Y_i)^{2}}{N}} \,,\\
 & \mathrm{Pearson\,correlation} && \mathrm{cor}(X,Y) = \frac{\sum_{i=1}^{N} (X_i - \mathrm{m}(X))(Y_i - \mathrm{m}(Y))}{\sqrt{\sum_{i=1}^{N} (X_i - \mathrm{m}(X))^2 \sum_{i=1}^{N} (Y_i - \mathrm{m}(Y))^2}} \,.
\end{empheq}

The formulas are adjusted in the case of circular variables to take into account their periodicity. Below the time series $X$ and $Y$ are assumed to be in radians between $-\pi$ and $+\pi$. For the standard deviation the formula given below is not in radians: the value ranges from $0$ to $\infty$ with $0$ describing a distribution with no variance at all. The formulas are:

\begin{empheq}{align}
 & \mathrm{mean} && \mathrm{m}_\mathrm{r}(X) = \mathrm{atan}\left(\mathrm{m}(\sin(X)),\mathrm{m}(\cos(X))\right) \,,\\
 & \mathrm{standard\,deviation} &&  \sigma_\mathrm{r}(X) = \sqrt{-2 \ln{\left(\sqrt{\mathrm{m}(\cos{X})^2 + \mathrm{m}(\sin{X})^2}\right)}} \,,\\
 & \mathrm{root\text{-}mean\text{-}square\,error} && \mathrm{rmse}_\mathrm{r}(X,Y) = \sqrt{\frac{\sum_{i=1}^{N} ((X_i - Y_i + \pi \mod{2\pi}) - \pi)^2}{N}} \,,\\
 & \mathrm{Pearson\,correlation} && \mathrm{cor}_\mathrm{r}(X,Y) = \frac{\sum_{i=1}^{N} \sin(X_i - \mathrm{m}_\mathrm{r}(X)) \sin(Y_i - \mathrm{m}_\mathrm{r}(Y))}{\sqrt{\sum_{i=1}^{N} \sin^2(X_i - \mathrm{m}_\mathrm{r}(X)) \sum_{i=1}^{N} \sin^2(Y_i - \mathrm{m}_\mathrm{r}(Y))}} \,.
\end{empheq}

\subsection{Validation against stations}

\begin{figure}
  \centering \includegraphics[width=10cm]{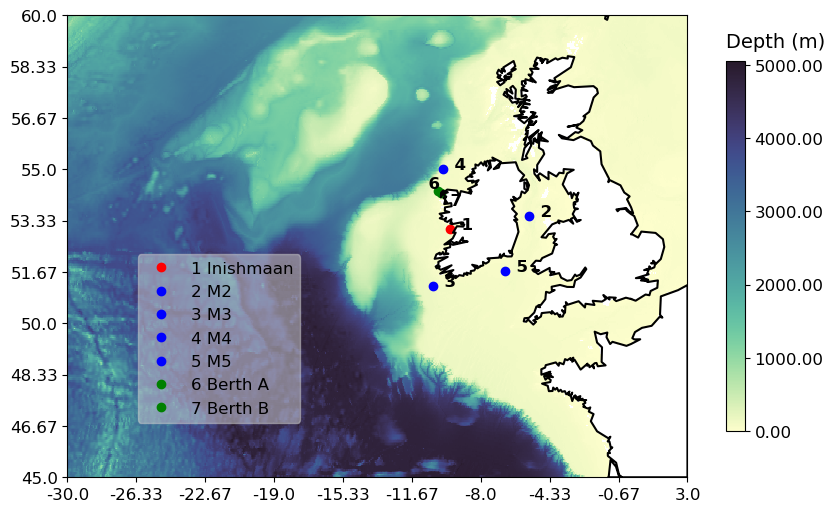}
  \caption{Location of the Inishmaan ADCP (red), wave buoys (green) and weather buoys (blue) used to validate the model. All the stations are located on the Irish shelf, in relative shallow waters (below $100 \,\mathrm{m}$.)}
  \label{stations}
\end{figure}

We first compare the output of the model with in-situ observations. The locations of the Acoustic Doppler Current Profiler (ADCP) and buoys are shown in figure \ref{stations}. They consist in one ADCP (Inishmaan), two wave buoys (Amets Berth A and Amets Berth B) and four weather buoys (M2, M3, M4 and M5). The ADCP was deployed near Inishmaan at the beginning of the year $2017$ with a $2 \,\mathrm{Hz}$ sampling rate. The original motivation for the deployment was to record and analyse high frequency time series for the surface elevation (\cite{fedele2019large}). The ADCP raw data is first processed into $12 \,\mathrm{min}$ time-averaged spectrum time series, and then into the desired mean wave parameters. The two wave buoys Amets Berth A and Amets Berth B are part of the Irish Wave Buoys Network. They measure a wide range of mean wave parameters, amongst which the significant wave height, peak wave direction and mean period $T_{m02}$, available with a $30 \,\mathrm{min}$ sampling rate, are retained. The peak period is also available for those two buoys but the data is of lower quality than the mean period. For consistency the mean period is also retained for the ADCP near Inishmaan. The four remaining weather buoys belong to the Irish Weather Buoys Network. Only the significant wave height, mean wave period $T_{m02}$ and mean wave direction are available from those four with a $30 \,\mathrm{min}$ sampling rate as well. All the directions are reported as the direction where the waves are coming from, with a clockwise rotation direction and the origin corresponds to waves coming from the North (nautical convention).

The instruments were not all recording constantly during the two years of hindcast. The model output data is truncated to match accordingly the observation. Spectral output is requested at the location of the buoys in the model every $10 \,\mathrm{min}$. The mean wave parameters of interest are then computed from the spectral output to match the exact quantity recorded by the station. To this effect the relative spectra directly computed by \ww{} at each station are transformed into absolute spectra using the relation (\ref{doppler}). In our case and in order to reduce the amount of variables to include in the analysis, it was decided to focus only on the significant wave height, available at all the stations, on the mean wave period and on the peak direction when available. If not then the mean direction is used. The wave systems are rarely composed of crossing swells in this region, but the mean direction or mean period can lose their physical meaning in the case of a combined wind sea and swell system, making it theoretically more consistent to rely on the peak direction and peak period when possible. The problem is well posed by \cite{portilla2015wave}, where a wave partitioning method is proposed before deriving the integrated parameters. A direct application of a crossing swell/wind sea problem is shown for instance by \cite{breivik2020combined} when computing the Stokes drift in such a situation.

\begin{figure}
  \centering \includegraphics[width=13cm]{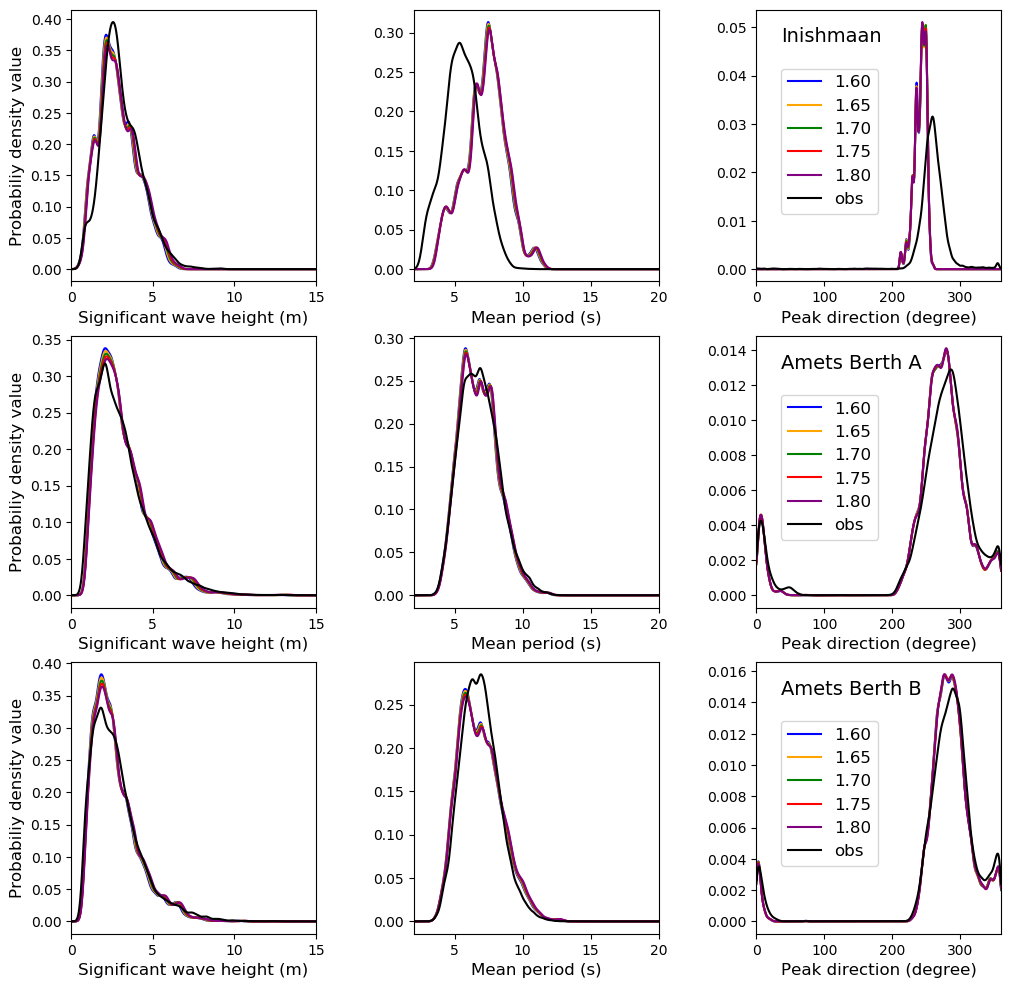}
  \caption{Probability density functions of the significant wave height (left column), mean period (middle column) and peak direction (right column), for the ADCP at Inishmaan and the two wave buoys Amets Berth A and Amets Berth B, and for the five different values of $\beta_{max}$ tested. There is a good visual agreement for all parameters for the Amets Berth A and Amets Berth B stations. For the Inishmaan ADCP, the mean period and peak direction are badly captured by the model, because nearshore effects are probably too strong. The influence of $\beta_{max}$ is visually appearing as small, but consistent between the stations.}
  \label{buoys-pdf1}
\end{figure}
\begin{figure}
  \centering \includegraphics[width=13cm]{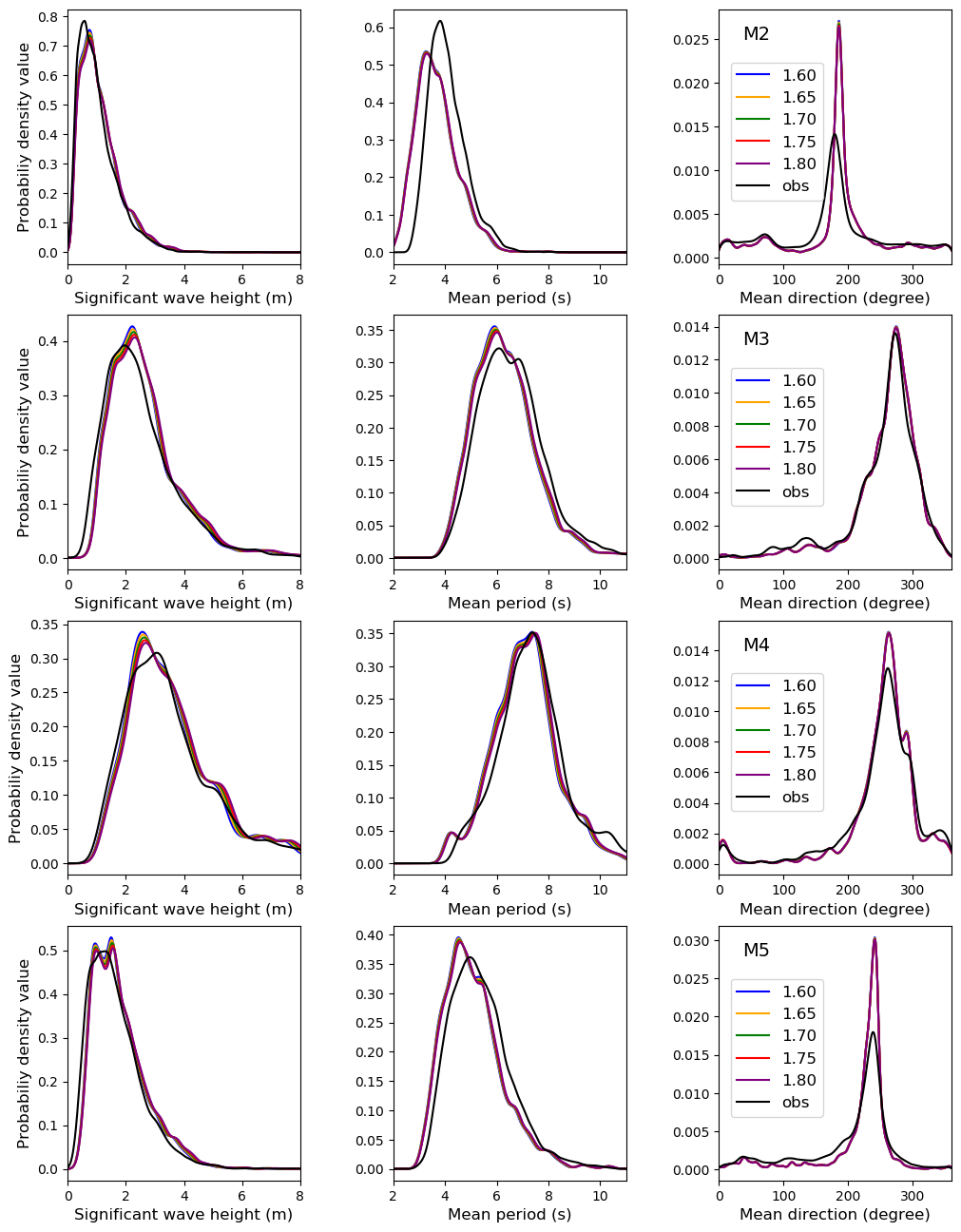}
  \caption{Probability density functions of the significant wave height (left column), mean period (middle column) and mean direction (right column), for the four weather stations used in the validation and for the five different values of $\beta_{max}$ tested. A good agreement is found overall for all the stations except for M2, located in the Irish Sea where the model is not expected to perform well.}
  \label{buoys-pdf2}
\end{figure}

We are first looking at the distribution of the different variables observed for the validation, shown in figure \ref{buoys-pdf1} and \ref{buoys-pdf2}, and how the growth rate is impacting their shape. Globally a good agreement is found between the model and the observations despite some buoys being located in nearshore areas where the model is not expected to perform well. More specifically, the significant wave height distribution agrees well with the observations at all locations, although a small bias is sometimes noticed for the position of the peak or its value. The same global comment goes for the mean period except for the ADCP at Inishmaan and the M2 buoy where a significant bias in the position of the peak is observed (respectively $+2.3 \,\mathrm{s}$ and $-0.5\,\mathrm{s}$). As for the wave direction the distribution is also well predicted except for the ADCP at Inishmaan where the model is showing a strong bias of $-15^o$. The model has also a tendency to narrow the distribution of the direction, predicting less variability in the direction of the waves (Inishmaan, M2, M5).

Looking at the impact of the growth parameter $\beta_{max}$, the differences induced on the shape of the probability density functions are slim, making it difficult to state anything from this analysis. The peak or mean direction is not impacted at all by the growth parameter, and the mean period only marginally. It is consistent with the way the $\beta_{max}$ parameter is impacting the model, acting homogeneously on all directions and frequencies. A small but noticeable impact is found on the significant wave height as a higher growth rate flattens the distribution by increasing the density of higher wave events. The impact is not a linear shift of the distribution as one could expect from the formulation (\ref{input1}). This is due to the non-linear terms appearing in the wave action and, although we are only modifying the input of energy in the system, the redistribution and dissipation are indirectly modified. Increasing the growth rate seems to improve the agreement for some stations (Amets Berth A, Amets Berth B, M4 and M5) and deteriorate it for others (Inishmaan and M2). For M3, it is hard to conclude as the peak value improves but at the same time its location deteriorates.

\begin{figure}
  \centering \includegraphics[width=14cm]{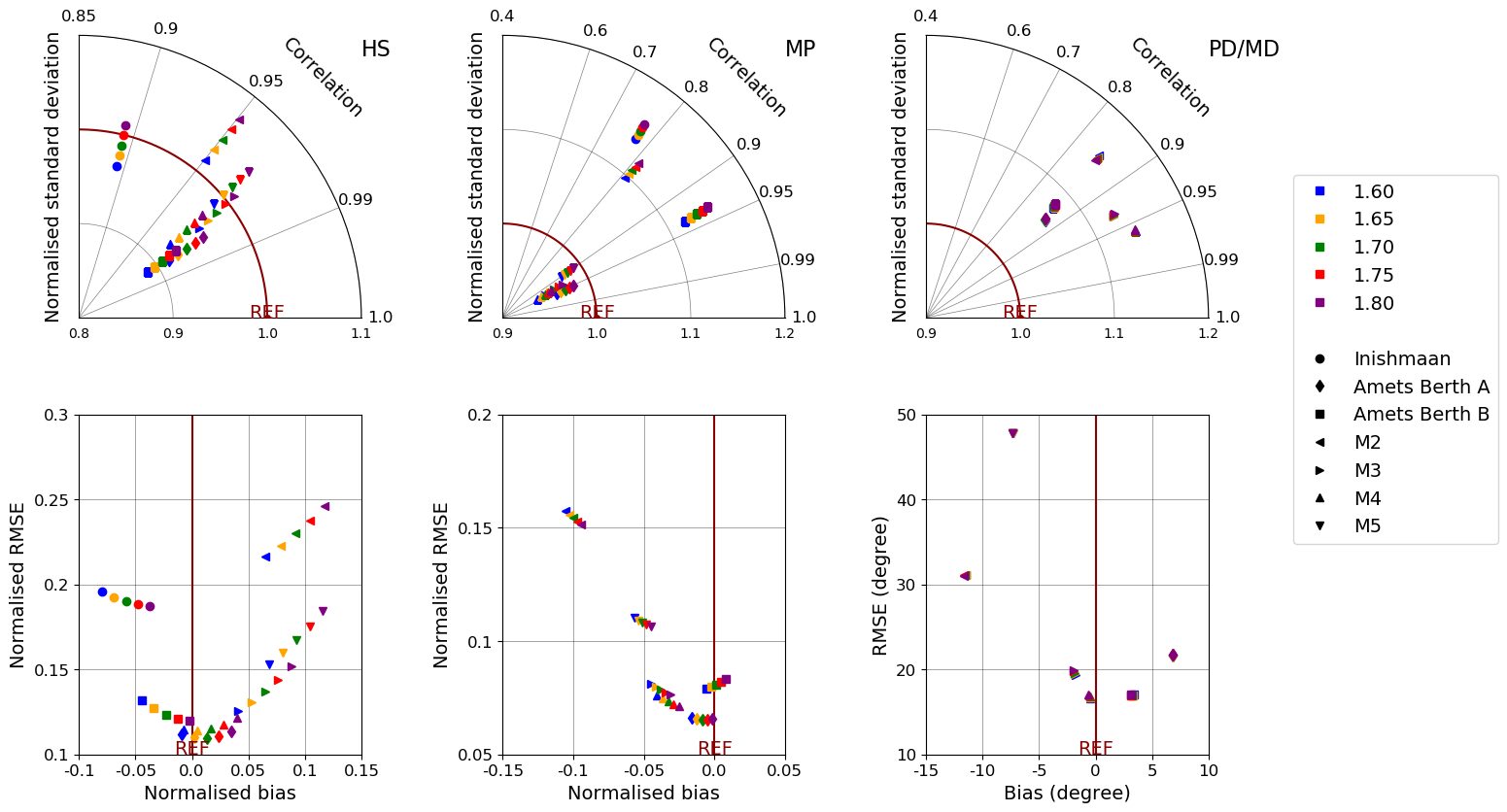}
  \caption{Taylor and RMSE versus bias diagrams for the significant wave height (left), peak or mean period (middle) and peak or mean direction (right), for the seven stations used in the model and for the five different values of $\beta_{max}$ tested. The value of $\beta_{max}$ has no impact on the correlation, only on the bias, RMSE, and standard deviation. The wave parameter the most impacted is the significant wave height.}
  \label{taylor-beta}
\end{figure}

In order to quantify the differences observed in the distributions, Taylor diagrams are shown in figure \ref{taylor-beta}, completed by RMSE versus bias diagrams, which are a good addition to the compact synthesis offered by the Taylor diagrams for which a constant bias between the data is not picked up. The probability density function plots give a broader view of the whole performance but they compare the distributions without checking that the events in the model and in the observation record occur simultaneously. As such they can over-estimate the agreement. This seems to be the case for the ADCP at Inishmaan and the two weather buoys M5 and M2 where the errors in direction are significant. For the first two, the respective error in standard deviation was too strong to be plotted on the Taylor diagram without ruining the scale.
Overall the same comment can be made that the growth rate impacts noticeably the significant wave height, but only marginally the mean period and not at all the peak or mean direction. Therefore only the significant wave height is retained for the analysis. A recurrent trend is observed with the bias consistently increasing with the growth rate for all stations, as the mean value from the model increases with the growth rate. The standard deviation also increases with the growth rate for all stations, meaning a higher growth rate puts more variability in the model as the distribution flattens and more high wave values are enabled. Surprisingly the correlation is consistently not impacted by the growth rate, and it can be clearly seen on the RMSE versus bias diagram that, depending on the stations, increasing the growth rate improves or not the performance. To be more precise, stations with a negative bias get a better agreement with a higher growth rate increasing the mean value and reducing the RMSE, whereas stations with a positive bias see their agreement deteriorated with a higher growth rate.

It transpires that using stations for the validation does not provide consistent insights about the behavior of the model. This is probably due to the coarse resolution and the nearshore locations of all the stations. It is likely that the errors introduced by the grid resolution on the bathymetry, wind input and propagation scheme are too important at this level and out-range the correction induced by the calibration of the wind input formulation.

\subsection{Validation against altimetry data}

We now compare the output of the model with satellite altimeter observations. This is more suited for large scale models as the observation is spatially averaging the area covered by the satellite beam, smoothing out at the same time the variability associated with small scale processes. The altimeter data used is provided by the Ifremer Laboratoire d'Océanographie Spatiale (LOS) and publicly available on their servers. Significant wave height measurements from nine altimeter missions are extracted, corrected and merged together (\cite{queffeulou2013merged}). A sub-square of the North-East Atlantic grid used in our model is used for the validation, also limiting the longitude at $-9^\circ{}$, excluding shallow water points where the bathymetry in particular induces large errors.
A satellite crosses the area of interest in a few minutes and because of memory space constraints the model output is requested hourly. Therefore two or three time steps appear in the time interpolation during the same sweep of a satellite, which is not an issue but is worth mentioning.

\begin{figure}
  \centering \includegraphics[width=14cm]{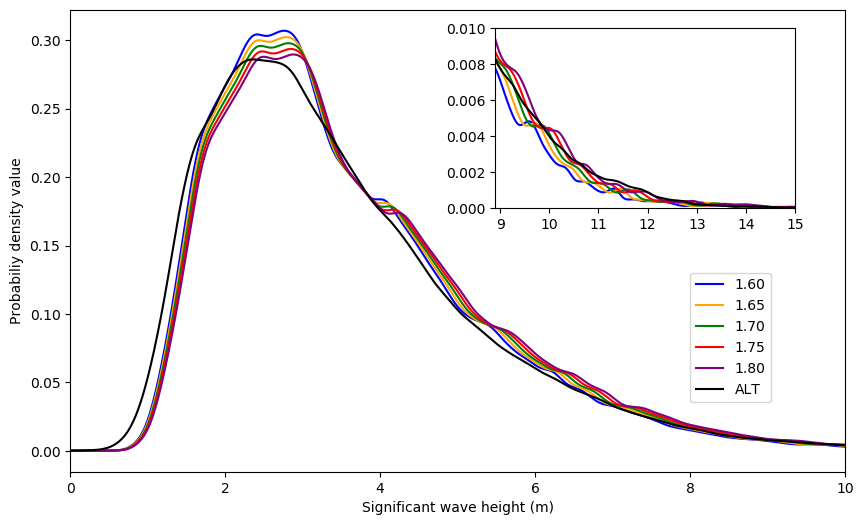}
  \caption{Probability density functions of the significant wave height for the five different values of $\beta_{max}$ tested (model against altimeter data). A zoom on the distribution tail is shown in the upper right. Higher values of $\beta_{max}$ give a distribution that is more spread, thus increasing the occurrence of higher wave events and reducing the occurrence of lower wave events. The smallest wave events, below $2 \,\mathrm{m}$, are badly captured by the model in all cases with little impact of $\beta_{max}$.}
  \label{pdf-alti}
\end{figure}

\begin{figure}
  \centering \includegraphics[width=14cm]{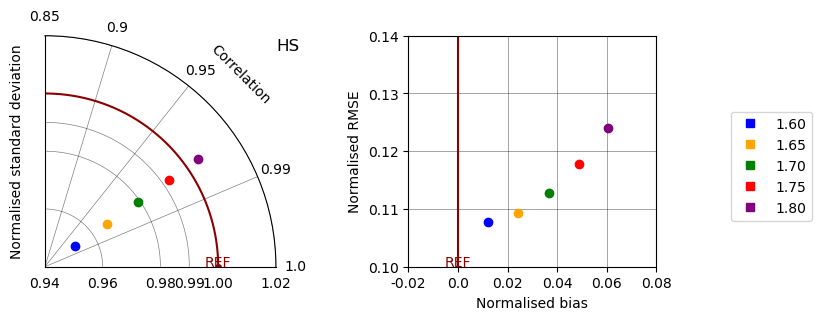}
  \caption{Taylor and RMSE versus bias diagrams for the significant wave height against satellite altimeter data, for the five different values of $\beta_{max}$ tested. The correlation is not impacted, an optimum could be inferred from the reduction in RMSE but that would fit the model for the most likely events characterised with small wave heights.}
  \label{taylor-alti}
\end{figure}

\begin{figure}
  \centering \includegraphics[width=12cm]{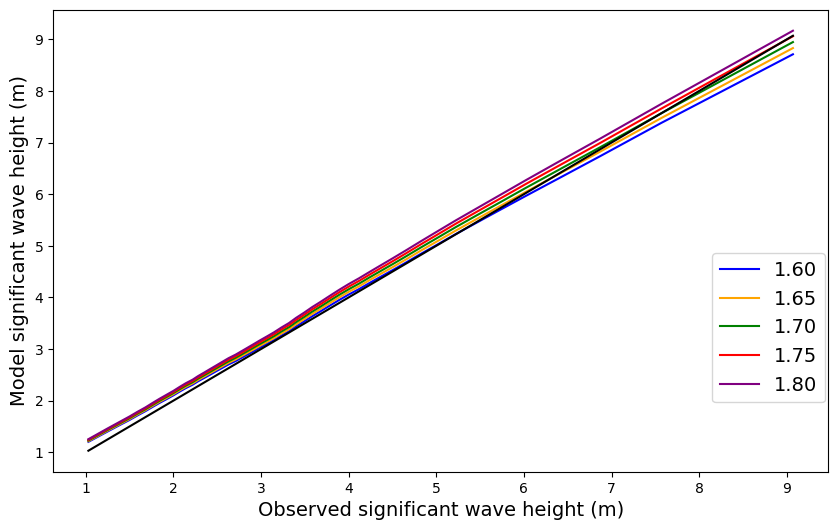}
  \caption{Q-Q plots of the significant wave height for the five different values of $\beta_{max}$ tested (altimeter data on the horizontal axis and model data on the vertical axis). A clear bias appears for the waves below $2\,\mathrm{m}$. The tail of the plot is used to infer an optimal $\beta_{max}$ value, fitting the model for the extreme wave events.}
  \label{qq-alti}
\end{figure}

The distributions on figure \ref{pdf-alti} are similar to what was observed for the different stations, showing very good agreement between the model and the altimetry data overall. The higher the wave height the lower the probability of encountering this event, with the last $1\%$ of the distribution corresponding to events with a significant wave height above 9 m.
The peak value is found around $2.4 \,\mathrm{m}$ and is well-captured by the simulation in general, although it is sensitive to the $\beta_{max}$ value.
The smaller the $\beta_{max}$ value, the more the distribution is shifted towards lower values and the narrower the distribution is around the peak position. As the growth rate increases, so does the occurrence of higher wave height events, showing a broader distribution with the peak position shifted more towards higher values.

The model consistently underestimates the small events whatever the value for $\beta_{max}$. The reason why it is so was not investigated thoroughly but it is believed to be correlated with the area chosen to compute the distributions. The initial growth of the distribution is observed around $0.5 \,\mathrm{m}$, while the model captures it around $0.8 \,\mathrm{m}$. A similar comparison done on the whole Atlantic domain, not shown in this paper, gives a better agreement for those extremely low wave height events, closer to what is observed in \cite{stopa2018wind} for instance. Some improvement could be made by adjusting the initial growth of waves for this area specifically, but it is not known whether the initial growth parametrization or low wind data is at fault here.

The observations based on the probability density are consistent with the Taylor diagram and bias plot shown in figure \ref{taylor-alti}. It can also be shown on the Taylor plot that the correlation is only marginally impacted by the growth rate, which can be the signature for a strong difference in bias solely, or for events with a low occurrence like extreme wave heights. The standard deviation is however significantly impacted. A value of $\beta_{max}$ between $1.75$ and $1.80$ seems to capture more accurately the spread of the distribution. Lower values for the growth rate are narrowing excessively the distribution. However for a value of $1.75$ the model is showing a positive bias of $5\%$ with a normalised RMSE of $12\%$. An optimal value of $\beta_{max}$ to minimise the bias would be reached for $1.60$ or even $1.55$ (extrapolating), and still giving a normalised RMSE of $11\%$, which corresponds to $0.39 \,\mathrm{m}$ for a mean value of $3.5$ m. From this analysis the growth rate seems to impact the model mostly in two ways, acting both on the bias and on the dispersion of the distribution, and there is no absolute optimal value that would optimise both parameters.

Using the bias to deduce an optimal value for the growth rate is not believed to be a good strategy. The probability density plot on figure \ref{pdf-alti} and the Q-Q plot on figure \ref{qq-alti} both show that in fact the low bias is obtained by balancing a positive bias for low wave heights with a negative bias for higher wave heights. The Q-Q plot also shows more clearly that none of the runs are capturing correctly the small wave events as they overestimate the wave heights of young wave systems.
Little difference is observed between the runs at this level. We thus conclude that the growth rate $\beta_{max}$ does not impact the initial growth rate of the waves. On the other hand the larger the wave height the more sensitive the coefficient becomes, with a global tendency of the model to underestimate the more extreme events. This non-linear impact explains the discrepancy between the bias and the normalised RMSE.

Increasing the growth rate value enables to improve the agreement for extreme events. An optimal value is found for $1.75$, which also gives the best agreement for the standard deviation. We decided to use this value in our model. A similar strategy was pursued by \cite{stopa2018wind}, who also makes the argument that extreme events are more harmful and more sensitive when it comes to coastal hazards, impacts on the communities and coastal infrastructure dimensioning.

\subsection{Comparison with existing work}

The accuracy of our model is compared with a relevant existing wave hindcast (\cite{gallagher2014long}) also focusing on Ireland. This $34$-year wave hindcast was performed using the unstructured version of \ww{} (version 4.11), forced by the \eraint{} winds. Several impactful differences between the model used in \cite{gallagher2014long} (hereinafter GA2014) and our model can be noted. An unstructured grid is used in GA2014 allowing to refine the nearshore region down to $250 \,\mathrm{m}$ when the most refined regular grid in our model has an average resolution of $3$ km. On the other hand GA2014 relies on wave spectra boundary conditions, taken from \eraint{}, when our model does not require boundary conditions at all. The wave hindcast done in GA2014 encompasses $34$ years ($1979$-$2012$) when our own analysis is performed over only two years ($2016$-$2017$). Therefore some variations in the wave statistics are to be expected regardless of the accuracy of each model, as proved by the mean observed values during each period shown in table \ref{sarah-buoys}. One important aspect that was not studied is the computational resources needed for each model. For information there are $40$ times more nodes in GA2014 than computational points in our model.

Despite the differences between the two models, a good agreement is achieved overall with both models showing comparable wave parameters - see table \ref{sarah-buoys}. For all stations the significant wave height and mean period predicted by our model show a slightly better agreement with the buoys than GA2014, with a smaller RMSE and higher correlation coefficient. The bias in each model is not consistent. For some stations and some variables, GA2014 can overestimate the wave parameter and our model underestimate it, and the other way around for a different station. As mentioned previously this can be explained by the complexity of the bathymetry impacting the propagation of swell systems or by the accuracy of the wind data impacting the local generation of wind sea.
None of the buoys are located in complex coastal features. Amets Berth A, Amets Berth B, M3 and M4 are facing the open ocean. M2 and M5 are respectively in the Irish sea and Celtic sea, which are more protected areas but still in deep enough water so that our model is still covering the location. Moreover none of the buoys are located in shallow water where the depth induced dissipation or wave breaking would dominate. These processes are better captured by a coastal model like GA2014. The four weather buoys M2, M3, M4 and M5 are all located far enough from the coast so that the resolution of the unstructured grid used in GA2014 is still coarse, similar to that of our model.

\begin{table}
\begin{center}
\resizebox{\textwidth}{!}{%
\begin{tabular}{l l|l l l l|l l l l|l l l l}
          &       & \multicolumn{4}{l|}{Hs}  & \multicolumn{4}{l|}{Mean period} & \multicolumn{4}{l}{Mean direction} \\
  Buoy    & Ref   & X   & Bias & RMSE & R    & X   & Bias & RMSE & R            & X     & Bias  & RMSE  & R \\
          &       & (m) & (cm) & (cm) &      & (s) & (s)  & (s)  &              & (deg) & (deg) & (deg) &   \\
  \hline
  Amets Berth A & GA2014 & $2.87$ & $5$ & $38$ & $0.96$ & $6.7$ & $0.2$ & $0.6$ & $0.92$ & $292$ & $9$ & $20$ & $0.70$ \\
          & Here   & $3.07$ & $7$ & $34$ & $0.98$ & $6.9$ & $-0.04$ & $0.4$ & $0.95$ & $282$ & $7$ & $22$ & $0.87$ \\
  \hline
  Amets Berth B & GA2014 & $2.77$ & $11$ & $40$ & $0.97$ & $7$ & $0.2$ & $0.7$ & $0.98$ & $296$ & $7$ & $16$ & $0.69$ \\
          & Here   & $2.78$ & $-3$ & $34$ & $0.98$ & $6.9$ & $0.03$ & $0.6$ & $0.93$ & $290$ & $3$ & $17$ & $0.86$ \\
  \hline
  M2      & GA2014 & $1.19$ & $15$ & $31$ & $0.94$ & $4.5$ & $0.9$ & $1.2$ & $0.65$ & $189$ & $-15$ & $24$ & $0.77$ \\
          & Here   & $1.03$ & $11$ & $24$ & $0.95$ & $4.1$ & $-0.4$ & $0.6$ & $0.81$ & $190$ & $-12$ & $31$ & $0.85$ \\
  \hline
  M3      & GA2014 & $2.86$ & $-4$ & $45$ & $0.95$ & $6.9$ & $0.3$ & $0.8$ & $0.87$ & $275$ & $5$ & $13$ & $ 0.95$ \\
          & Here   & $2.53$ & $19$ & $36$ & $0.97$ & $6.6$ & $-0.2$ & $0.5$ & $0.93$ & $269$ & $-2$ & $20$ & $0.93$ \\
  \hline
  M4      & GA2014 & $3.11$ & $-1$ & $39$ & $0.97$ & $7$ & $0.2$ & $0.7$ & $0.98$ & $275$ & $2$ & $13$ & $0.94$ \\
          & Here   & $3.57$ & $10$ & $42$ & $0.97$ & $7.4$ & $-0.2$ & $0.5$ & $0.93$ & $265$ & $-1$ & $17$ & $0.96$ \\
  \hline
  M5      & GA2014 & $1.81$ & $-3$ & $38$ & $0.94$ & $5.5$ & $0.1$ & $0.8$ & $0.82$ & $231$ & $-6$ & $18$ & $0.84$ \\
          & Here   & $1.65$ & $17$ & $29$ & $0.97$ & $5.4$ & $-0.3$ & $0.6$ & $0.90$ & $233$ & $7$ & $48$ & $0.51$ \\
  \hline
\end{tabular}}
\caption{Wave statistics for the different buoys used to validate the model. The results from \cite{gallagher2014long} are compared with the model used in this paper. The peak direction is used for the two wave buoys Amets Berth A and Amets Berth B. $X$ corresponds to the mean value recorded by the buoy. Overall similar if not better statistics appear for the model used in this paper.}
\label{sarah-buoys}
\end{center}
\end{table}

As for altimetry data, with the significant wave height statistics shown in table \ref{sarah-alti}, a better agreement is obtained for GA2014 with lower bias and lower RMSE, for a similar correlation coefficient. It is however ambiguous to conclude anything from those statistics as for both models they are computed differently. The sub-region used in GA2014 includes the Atlantic and Celtic Sea, whereas in our case only the Atlantic region is included, meaning that we are not comparing at the exact same locations. The difference in the recorded significant wave height is the proof of that, with $2.69 \,\mathrm{m}$ for the analysis conducted in GA2014 and $4.17 \,\mathrm{m}$ in our case. The area covered in GA2014 includes more shallower and protected areas, where obtaining a good agreement is more difficult. GA2014 seems to outperform our own model.

\begin{table}
\begin{center}
\begin{tabular}{l|l|l l l l}
   Ref & Region & Mean observed Hs (m) & Bias (cm) & RMSE (cm) & R    \\
  \hline
   GA2014 & Atlantic and Celtic Sea & $2.69$ & $5$  & $39$ & $0.97$ \\
   Here   & North-East Atlantic     & $4.17$ & $19$ & $46$ & $0.97$ \\
  \hline
\end{tabular}
\caption{Computed significant wave height statistics against satellite altimeter observations. The results from \cite{gallagher2014long} are compared with the model used in this paper. Note that the domains are different, hence the difference in mean values, bias and RMSE. However the correlation coefficients are similar.}
\label{sarah-alti}
\end{center}
\end{table}

In conclusion we are satisfied with the performance of our \ww{} model. It seems to predict accurately the swell systems propagating to Ireland with an accuracy matching that of a coastal model like GA2014.
However the discrepancy in peak or mean direction suggests that the nearshore processes are poorly captured, which is not a surprise given the resolution and the complex bathymetry of the Irish shelf showing high localised gradients. It is also reasonable to assume that the locally generated wind sea is badly represented in the model, as highlighted by the negative bias for young and small waves in figure \ref{pdf-alti}, but it does not seem to impact too much the accuracy of the model at the locations of the stations.

\section{Impact of the current field}

The \ww{} model validated in the previous section is now used to evaluate the impact of the current field on the accuracy of the model. The two-year hindcast is compared against a reference solution without current interactions.

\subsection{Global results}

\begin{figure}
  \centering \includegraphics[width=14cm]{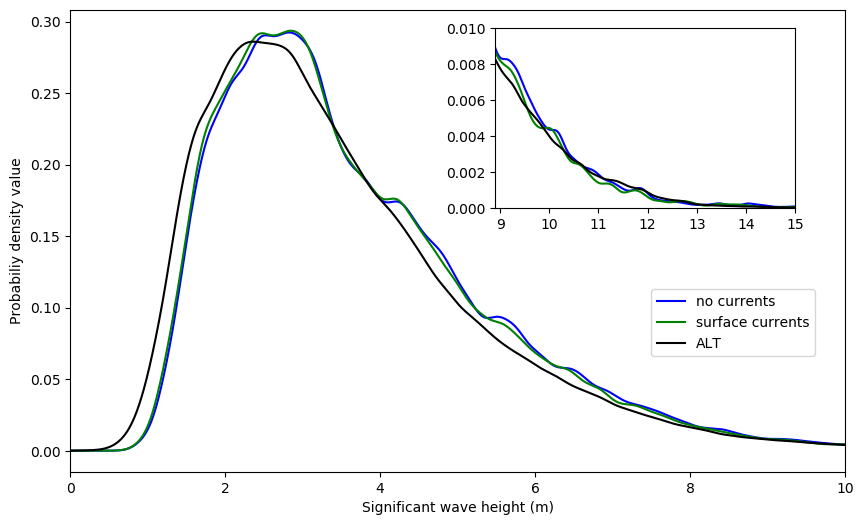}
  \caption{Probability density functions of the significant wave height, comparing the run with surface currents and the run without current interactions against altimetry data. A zoom on the distribution tail is shown in the upper right. The impact of currents is small, and no specific trend is noticeable.}
  \label{pdf-alti-cur}
\end{figure}

\begin{figure}
  \centering \includegraphics[width=14cm]{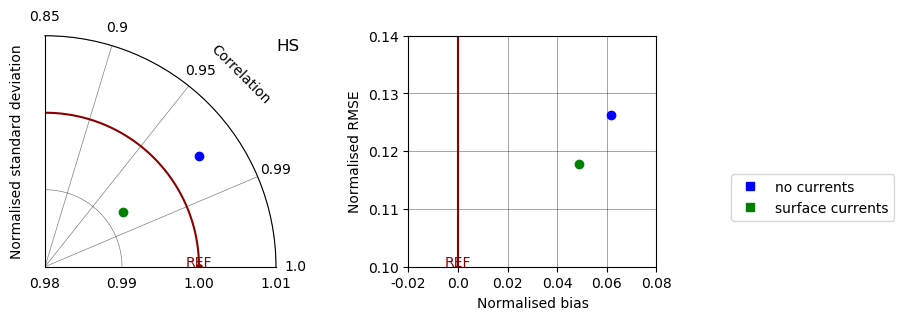}
  \caption{Taylor and RMSE versus bias diagrams for the significant wave height against satellite altimetry data, comparing the run with surface currents and the run without current interactions. Differences are of the order of $1\%$ for the standard deviation, bias and RMSE. No change in correlation is observed. The run with current interactions gives a better agreement.}
  \label{taylor-alti-cur}
\end{figure}

We start with a global overview comparing the two runs against altimetry data. The probability density functions shown on figure \ref{pdf-alti-cur} are very similar with only marginal and point-wise differences. No trend appears, even for extremely low or high events.
Looking at the statistics in figure \ref{taylor-alti-cur}, the correlation coefficient is not impacted at all, and the standard deviation only marginally with a $1\%$ difference.
The differences in terms of normalised bias and RMSE are also very small, around $1\%$ for both the bias and RMSE. The solution without currents gives the worst agreement with a higher bias and RMSE.

\begin{figure}
  \centering \includegraphics[width=14cm]{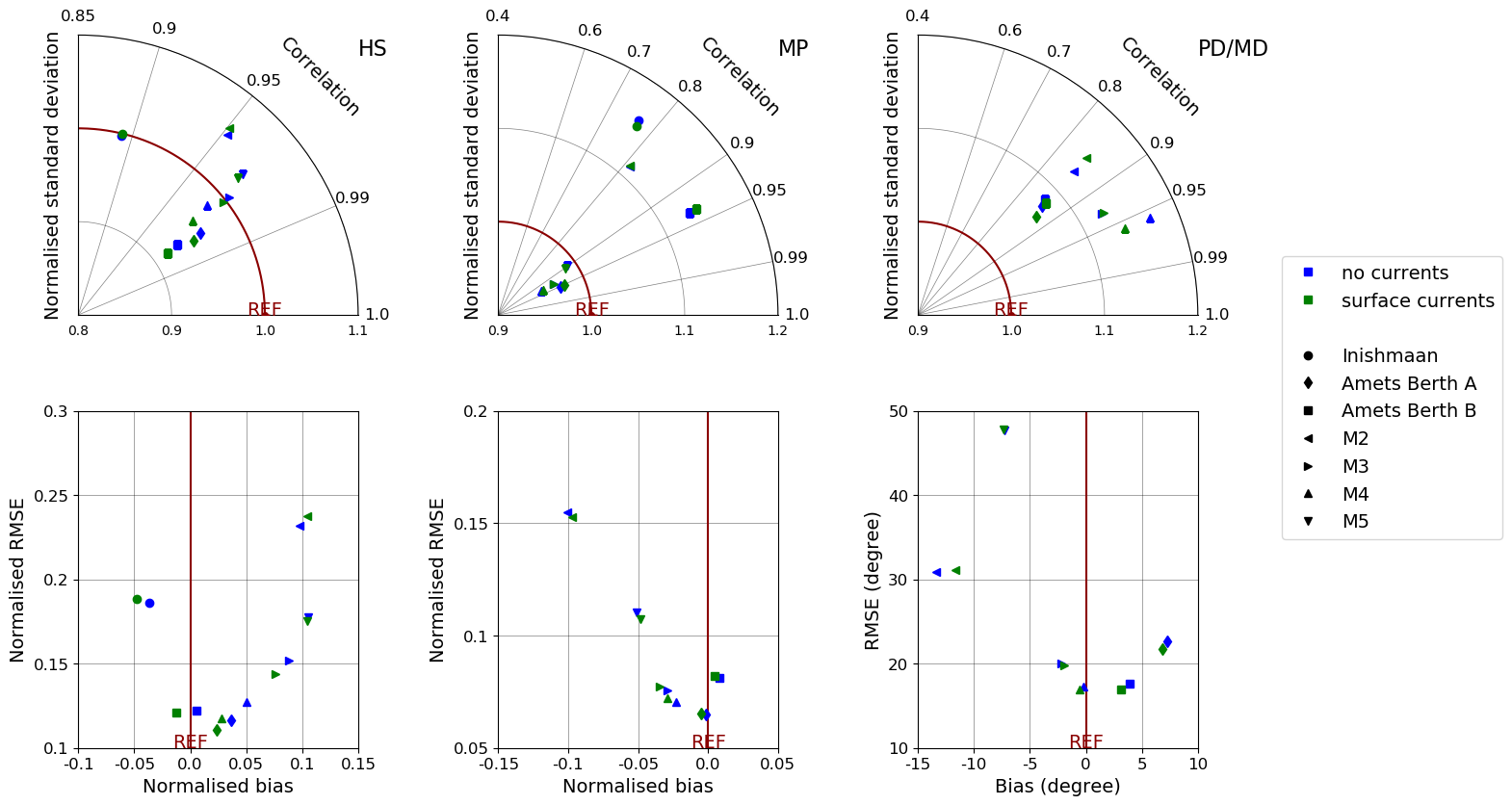}
  \caption{Taylor and RMSE versus bias diagrams for the significant wave height (left), peak or mean period (middle) and peak or mean direction (right), for the seven stations used in the model, comparing the run with surface currents and the run without current interactions. Currents do not always improve the agreement, and the impact of currents on the statistics is marginal compared with the spread between all the stations.}
  \label{taylor-buoys-cur}
\end{figure}

Despite the comments given in the previous section about using buoys for such a large scale simulation we show some of those statistics in figure \ref{taylor-buoys-cur}. This is motivated by the idea that current induced refraction is a local process but the focusing or dispersion of the waves can still be observed down-wave, as mentioned by \cite{ardhuin2012numerical}.
The significant wave height, the mean period and the wave direction overall show only marginal differences both in terms of distribution and in terms of biases. The impact of currents is of the order of $1\%$. Despite being almost marginal for some stations, the influence of currents is more noticeable with a clear separation between the run without currents and the run with currents (Amets Berth A, Amets Berth B, M2 and M4 notably). For Amets Berth A, Amets Berth B and M4 their location in an area known for having stronger currents can explain this observation. There is no consistency however in terms of statistics impacted. For instance the mean direction standard deviation at M4 shows a $3\%$ difference, but only a marginal difference in bias or RMSE for the mean direction, and no difference at all in terms of mean period.
It is also noted that the spread either in correlation coefficient, standard deviation error or bias between the stations themselves is more significant than the spreading induced by currents.

The impact of currents observed against either altimetry data or in situ buoys remains small, especially if compared to how sensitive the model is towards the wind input, which makes any conclusion hazardous to draw. There is however a tendency for currents to overall improve the accuracy of the model. Looking at the significant wave height only, the normalised standard deviation is also shown to be reduced due to the impact of currents. It slightly transpires on the distributions (figure \ref{pdf-alti-cur}), as currents seem to narrow the distribution around peak value.

\subsection{Spatial variability}

\begin{figure}
  \centering \includegraphics[width=14cm]{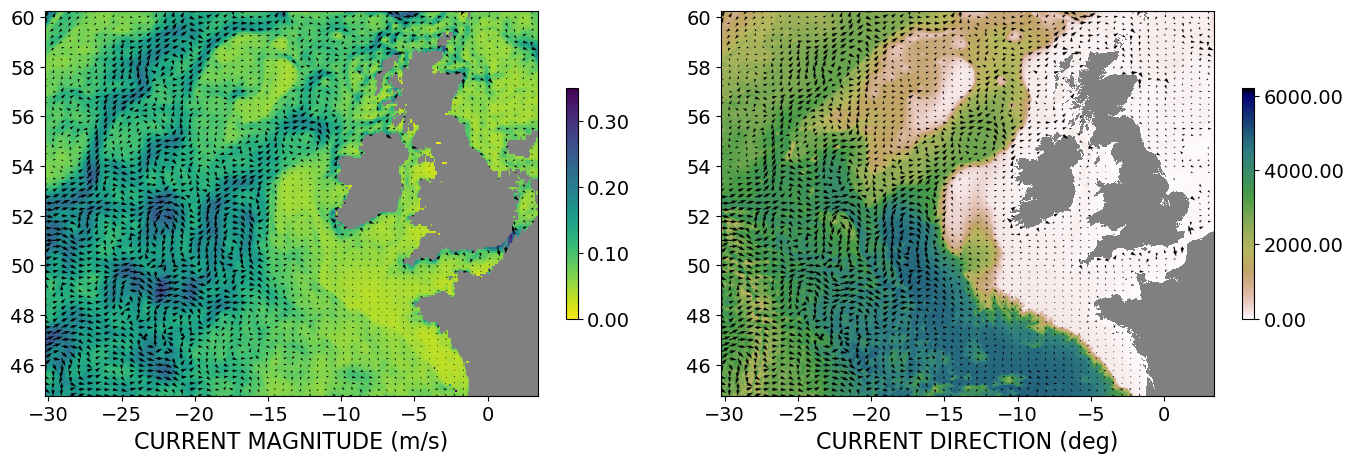}
  \caption{Mean average of the surface current magnitude and direction. The left plot shows the magnitude as a color map with vector arrows for the current. The right plot shows the bathymetry as a color with vector arrows for the current. Stable gyres appear outside the Irish shelf over the North Atlantic Current.}
  \label{cur-mean}
\end{figure}

Looking at the statistics does not give a lot of information on the process as to how and where exactly the small observed differences appear. In this section the spatial variation induced by the currents and the scale at which they impact the wave propagation are analysed for the present application. It was shown in \cite{ardhuin2017small} and \cite{marechal2020surface} that currents have an impact at short spatial scales, between 10 km and 100 km. These two studies focus on two strong boundary currents (Agulhas and Gulf stream respectively).
Our case features the North Atlantic Current but it is already weak and diffused in this region, as highlighted in figure \ref{cur-mean} that shows the two-year averaged current field. Stable geostrophic gyres still appear. They are eddies detaching from the main current mostly due to the interaction with the bathymetry.

\begin{figure}
  \centering \includegraphics[width=14cm]{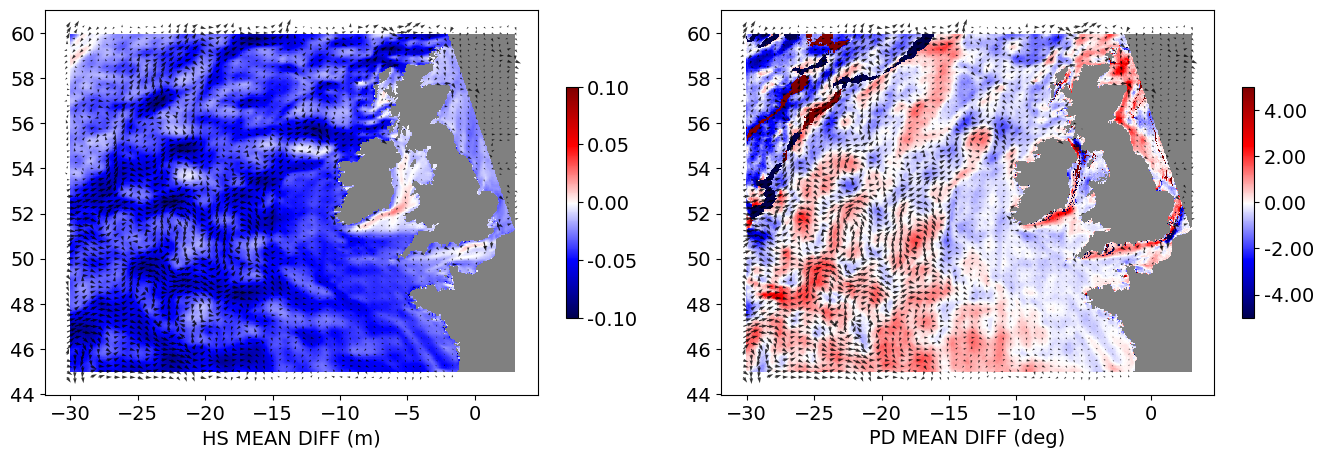}
  \caption{Significant wave height difference (left) and peak direction difference (right) between the solution with current interaction and the solution without currents. The vector arrows are the mean values of the current. A spatial correlation between theses biases and the oceanic gyres appears.}
  \label{wave-diff}
\end{figure}

There is a clear correlation between those mesoscale geostrophic structures and the propagation of waves, both in terms of significant wave height and direction, as shown in figure \ref{wave-diff}. The total average difference between the run with current interaction and the run without interaction is shown, both for the significant wave height and the peak direction. The impact of current is still small. The currents induce a variation of $10 \,\mathrm{cm}$ at best for the significant wave height and $5^o$ for the peak direction.
A stronger impact is observed in nearshore regions and narrow regions like the Channel passage and the Irish Sea. However, the resolutions of our model and of the \glorys{} model are too coarse to correctly capture nearshore processes so they are ignored. The peak direction is also showing some unexpected large differences in the North-West regions (see the darker red and blue areas in the left plot of figure \ref{wave-diff}). It corresponds to a somehow strong change in peak direction. The location of this strong shift is slightly different for the two runs. The existence of this front or the reasons why it is being affected by currents are not explained.
Weaker impacts of the current are observed for both the wave heights and directions on the Irish shelf. This is mostly due to currents being weaker on the shelf, as tides are not resolved in the \glorys{} product used. The two-year time average used to derive the statistics also probably cancels out some part of the impact as there is still some variability in the swell direction reaching this region. It would require additional investigation but one would still expect the refraction features to be spatially coherent to some extent and propagate down-wave, along with the swell towards the coast.

\begin{figure}
  \centering \includegraphics[width=8cm]{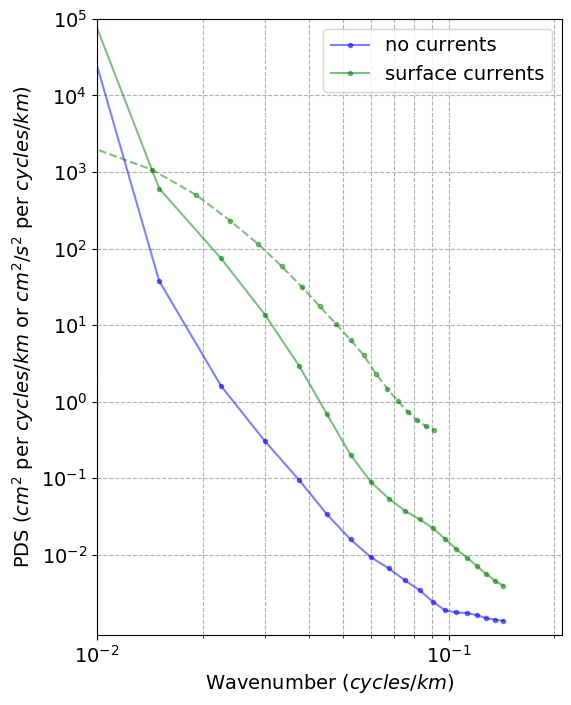}
  \caption{Power density spectra (PDS) of the currents and significant wave height with and without current interaction. Solid lines are for the significant wave height, in $\mathrm{cm}^2/(\mathrm{cycles}/\mathrm{km})$, and dashed lines are for the currents, in $(\mathrm{cm}^2/\mathrm{s})/(\mathrm{cycles}/\mathrm{km})$. The two-dimensional spectra are computed in a box contained between $-29^\circ{}$ and $-11^\circ{}$ longitude and $46^\circ{}$ and $59^\circ{}$ latitude. They are converted in one-dimensional spectra by averaging over the direction of the wave-vector. Currents impact and explain the wave field at short scales. We expect the PDS decay for the waves to match that of the currents but we are limited by the resolution of the model and \glorys{} product to realize this observation.}
  \label{psd}
\end{figure}

The one-dimensional power spectral density functions are shown in figure \ref{psd}. Following what is done in \cite{ardhuin2017small}, they are computed by averaging over time the two-dimensional power spectral density obtained with a Fourier transform of the currents and wave field, then averaging over the direction of the wave-vector to reduce it to one direction. The wavenumbers are expressed in cycles per kilometer, later written cpk in this paper.

Compared to \cite{ardhuin2017small} the currents in our case carry less energy overall, which is expected as the North Atlantic Current is weaker than the Gulf Stream. A nearly  constant slope is observed for the currents spectral density, which is expected from the Kolmogorov theory.
The resolution of the \glorys{} product does not allow to capture scales smaller than $10 \,\mathrm{km}$ ($k < 0.1 \,\mathrm{cpk}$). However, the wave model interpolates the current field on the computational grid, which explains the difference in slope for the significant wave heights at those smaller scales.

At scales of $100 \,\mathrm{km}$ and more ($k > 0.01 \,\mathrm{cpk}$) the currents show no impact. The spatial variability is most likely explained by the propagation of swell systems. For smaller scales the currents add more spatial variability in the wave model. In \cite{ardhuin2017small} the impact of currents was already noticeable at a $100 \,\mathrm{km}$ scale ($k = 0.01 \,\mathrm{cpk}$), whereas in our case the threshold is around $50 \,\mathrm{km}$ ($k = 0.02 \,\mathrm{cpk}$). The run with currents predicts $20$ times more energy at this scale making the currents themselves explain $95\%$ of the spatial variability at this scale.

Focusing on the scale around $50 \,\mathrm{km}$ ($k = 0.02 \,\mathrm{cpk}$), the effect of currents is to align the wave spectral density slope with the current spectral density slope. In \cite{ardhuin2017small} currents are stronger and those slopes are actually equal.

\subsection{Following tracks}

\begin{figure}
  \centering \includegraphics[width=14cm]{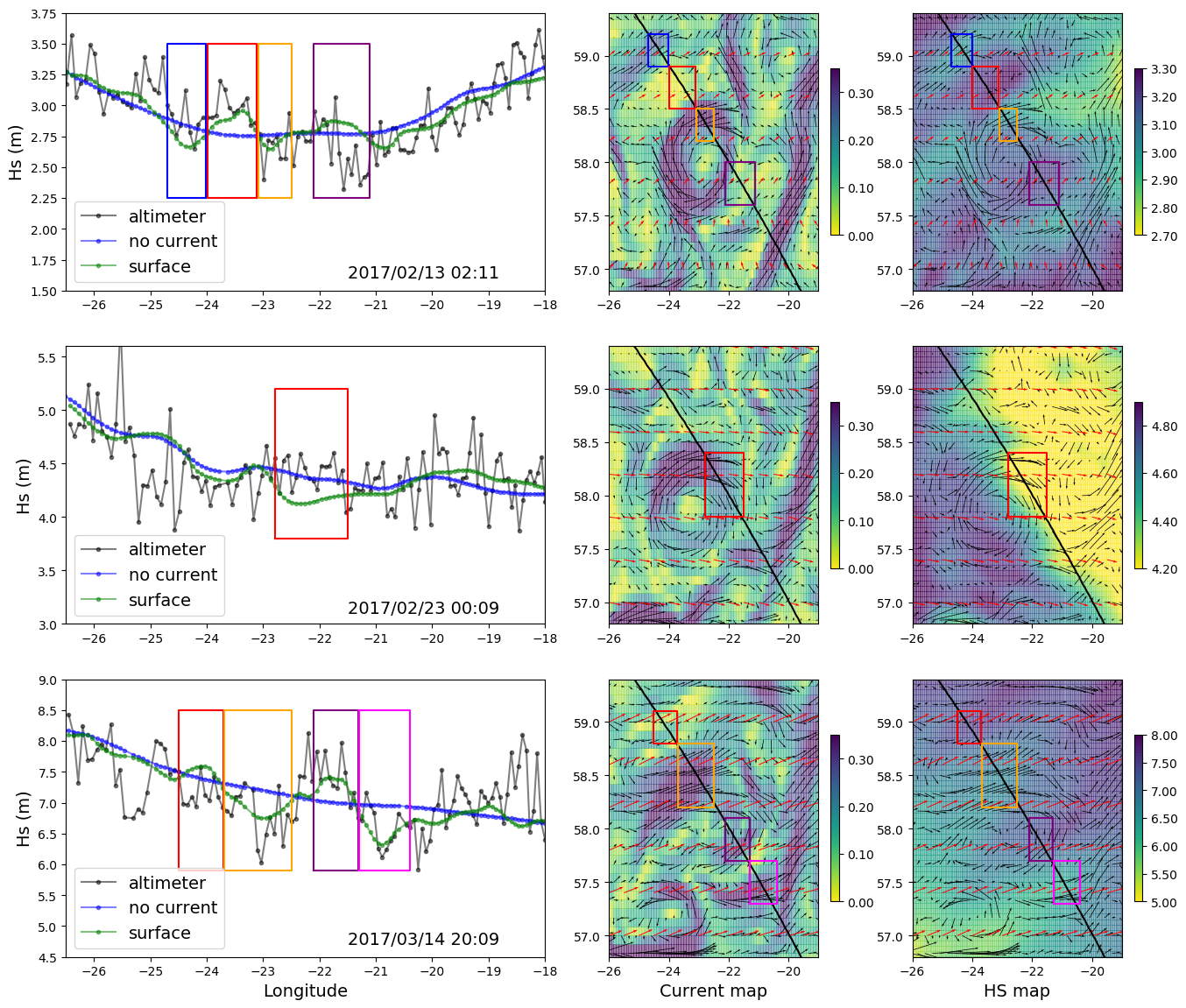}
  \caption{Following tracks altimetry data compared with model output. The same track at three different times is shown. Maps of the current (middle) and of the significant wave height (right) for the run with currents are also shown at the satellite track time. The black vector arrows represent the current and the red ones the mean wave direction. The impact of currents is not always seen to improve the agreement although some text-book situations are highlighted. The signal recorded by the altimeter is also showing a strong variability, which is matching the effects of currents, making it difficult to compare the data.}
  \label{tracks}
\end{figure}

In the previous section it was emphasized that the impact of currents is only visible at short spatial lengths. Both global and station time-averaged statistics were also shown to give little to no evidence about the benefits of enabling current interactions. We now look at concrete examples to see in more detail how the interaction appears in the model. One particular track of the satellite Jason2 that crosses a stationary stable geostrophic gyre was chosen for the example, at three different times, as shown in figure \ref{tracks}. The area chosen is off the coast, in deep enough water so that nearshore effects do not impact the quality of the altimetry data. The satellite only takes a few minutes to cover the area shown. Compared to the hourly output from the model the satellite data can safely be assumed to be a spatial cross-section observation of the area at a given time and not a time-series. All the model outputs are time-interpolated to match the time of satellite observation, and for the first time series (2017-02-13) the model output was added a $35 \,\mathrm{cm}$ negative bias to better match the observations. This was done to remove the errors possibly induced by the wind input and propagation of the swell systems, as only the impact of currents is of interest here, which appears as a signal perturbation at short scales.

A noticeable difference of the order of 20--50 cm can be observed between the run without current interaction and the run with current interaction. The agreement is not always better for the run with currents. Jason2 has an accuracy of $2 \,\mathrm{cm}$, the record frequency is $1 \,\mathrm{Hz}$ corresponding to $1 \,\mathrm{s}$ sampling period averaging an area of roughly $5 \,\mathrm{km}$ (varying with the significant wave height). The sampling frequency is way higher than the model output frequency of $1 \,\mathrm{h}$ and as a result additional wave components with lower time-scales are included in the record, bringing additional short scale variability in the data-set. This variability, looking at the data, is estimated to have an amplitude of the order of $20 \,\mathrm{cm}$, which is unfortunately comparable to the current induced variability.

A clear correlation between strong current gradients and locations of enhanced or reduced significant wave height is observed in figure \ref{tracks}, especially for the first row (2017-02-13) featuring a homogeneous wave field across the region. In the second and third rows the wave field is more heterogeneous. A swell system enters the area of interest for the second row (2017-02-23) whereas in the third row (2017-03-14) a swell system leaves the area. Although a more thorough analysis removing the trend induced by propagating swell would be more conclusive, it looks like in all cases the effect of currents is only contained within the zone of strong current gradients and no coherent shapes emerge down-wave of the refraction. This goes against what is observed in \cite{ardhuin2012numerical}, where current effects do have an impact down-wave. The continuous presence of current features and gradients prevents the propagation by inducing new refraction continuously, and the length scale at which those structures would be coherent might be too short to be observed with the resolution of our numerical model. It is also worthwhile mentioning that in \cite{ardhuin2012numerical} the observation is made in a much more coastal area featuring complex bathymetry, whereas in this case the waves are propagating in deep water.

Focusing now on the current induced effect, the first row shows a text-book example of current refraction. Looking at the first three boxes (blue, red and orange) following the satellite track, perpendicular to the wave propagation, the response of the model is coherent with what is expected from current refraction. The mean wave direction is aligned with the current direction. In the blue and orange side boxes the currents is way stronger than in the red center box. As a result the wave energy is being refracted in this same center box giving a higher wave height than in a case without current interaction. It also seems to agree more with the altimetry observation. The same process and good agreement is also observed for the four boxes in the third row.
However disagreement appears for the purple box in the first row. The run with current interaction predicts an increase in wave height when the satellite shows a decrease. The same goes for the red box in the second row, but this time currents in the wave model induce a decrease in wave height when the observation agrees well with a model without current interaction. In those two cases the situation is more complex as currents turn. It is not straightforward to apply the theory but it shows than current interaction can also induce errors in the wave propagation. The occurrences of those events were not studied.

\section{Conclusion}

In this paper we presented a \ww{} model forced with the \era{} winds and currents and sea level from \glorys{}. The purpose of this model is to generate accurate and high resolution swell conditions for Ireland. The model is described and consists of three nested grids. The wind input with the ST4 package is optimised for the specific model set-up in this paper. Disregarding nearshore wave and weather buoys, judged unsuitable for the resolution in the model, we used satellite data derived from altimetry measurements.
We found that the best growth rate value for the application shown in this paper is $\beta_{\mathrm{max}} = 1.75$, which is higher than the default value but still within the range of values found in the literature. The strategy that we followed was to give the best agreement possible for the extreme wave events, despite deteriorating the agreement for the most frequent wave systems. This strategy also enables the wave model to better capture the spread of the distribution.

Discrepancies in the distribution of wave heights are still found. The peak of the distribution is around $2.4 \,\mathrm{m}$. The wave model overestimates it by $0.3 \,\mathrm{m}$. Extremely low wave heights are badly picked up and underestimated by the model. The initial growth of the distribution is observed around $0.5 \,\mathrm{m}$, while the model captures it around $0.8 \,\mathrm{m}$. This error seems to be correlated with the area chosen to compute the distribution. A global comparison, not shown in this paper, gives a better agreement for those low wave height events. On the contrary, extreme wave height systems are in general underestimated and the model was optimised in order to correct for this negative bias.
A manual correction of the high winds was also conducted in order to correct this negative bias.

We used the model to investigate the effects of currents on the wave propagation over a large scale model, focusing on the region of Ireland which contains a portion of the North Atlantic Current generating mesoscale eddies on the approach of the Irish shelf. Comparing the model against altimetry data, it is found that taking into account the current interaction reduces the error by $1\%$. In agreement with the literature we also found that currents impact the wave field at very small scales and explain the majority of the wave height variability at those scales, up to $95\%$ below $50 \,\mathrm{km}$. Those processes occur at the limit of the scales that can be resolved by our model. The computational grid resolution is indeed $3 \,\mathrm{km}$, and $6 \,\mathrm{km}$ for \glorys{} which provides the current field. A strong correlation between the impact on the wave field and the presence of the mesoscale eddies is also found.

We studied in more detail the effects of currents on the wave field looking at particular snapshots. The model is found to capture accurately the refraction of waves induced by currents, but some unexpected and unexplained behaviors are observed as well where the model does not follow the behavior observed by the satellite. On the same topic the impact of currents on the significant wave height is found to be of the same order as the variability inherently present in altimetry derived observations, which is evaluated around $0.5 \,\mathrm{m}$. This is deemed to be an issue for objectively assessing the gain in accuracy provided by modelling the current interaction at this scale. A better treatment of the altimetry signal would be needed to allows for more relevant comparisons.

The last feature that was not mentioned nor studied is the impact of currents through the wind input term as the wind speed is corrected by the surface currents. This is still a trending topic and using a relative wind speed has been shown to improve the accuracy of the model (\cite{ardhuin2012numerical}, \cite{renault2016modulation}, \cite{ardhuin2017small}).

\section*{Acknowledgments}{The authors wish to acknowledge the Cullen Fellowship Programme, under the fellowship ``Coupled wave-ocean models'' funded by the Marine Institute. The authors also wish to acknowledge the DJEI/DES/SFI/HEA Irish Centre for High-End Computing (ICHEC) for the provision of computational facilities and support.}

\newpage
\printbibliography

\end{document}